# SEMANTIC WEB PREFETCHING USING SEMANTIC RELATEDNESS BETWEEN WEB PAGES

**Jyoti\* and Jyotsna Parmar\*\***

\*Department of Computer Science, YMCAUST Faridabad, India
\*\*Department of Computer Science, YMCAUST Faridabad, India

**ABSTRACT:** Semantic web prefetching makes use of the semantics of the keywords provided by the user for making predictions. These keywords can be descriptive texts like anchor texts, texts surrounding anchor texts, titles or simple text of the content. For finding the relation between two webpages, semantic information can be used. Semantic information is embedded within the web pages during their designing for the purpose of reflecting the relationship between the web pages. The client can fetch this information from the server. However, this technique involves load on web designers for adding external tags and on server for providing this information along with the desired page, which is not desirable. We are trying to find the semantic relation between web pages using the keywords provided by the user and the anchor texts of the hyperlinks on the present web page. Various semantic relations are: sequential (seq), similar (sim), cause-effective (ce), implication (imp), subtype (st), instance (ins), referential (ref). Their priorities are in the order: seq>sim>ce>imp>st>ins>ref. This paper presents algorithms for finding sequential and similar relations between web pages in order to improve the performance of web prefetching with no overhead on designers and load on server. These algorithms are to be implemented on the client side. These algorithms do not involve any overhead of web designers and extra load on server.

## INTRODUCTION

In today's digital age, when the internet has revolutionized the world by spreading at a pace that no one had imagined, internet has become a way of life for human beings. Though internet speed has increased, client is still facing challenge with the latency time or propagation time. To reduce latency, caching and prefetching techniques can be used. However, caching the web pages on the client system is not good for the dynamic websites which keeps on changing rapidly. Another technique is web prefetching, which pre-fetches and stores the webpages in prefetching cache that the user is likely to request for in the future. Semantic web prefetching makes use of keywords and descriptive texts like anchor text, titles, text surrounding anchor text of the present web pages for predicting user's future requests. To find relation between two web pages, semantic information can be used. Semantic information is embedded within the web pages during their designing for the purpose of reflecting the relationship between the web pages. The client can fetch this information from the server. However, this technique involves load on web designers for adding external tags and on server for providing this information along with the desired page, which is not desirable. This paper is an effort to find the semantic relation between web pages using the keywords provided by the user and the anchor texts of the hyperlinks on the present web page. It provides algorithms for sequential and similar semantic relations. These algorithms will get implemented on the client side which is not going to cause overhead on designers and load on server for semantic information.

The rest of the paper is organized as follows: section 2 is a literature survey of various prefetching techniques. Section 3 is the proposed work which contains the implementation of sequential and semantic association. Section 4 is the conclusion of this paper. Section 5 provides future work in these implementations and section 6 contains the references.

**LITERATURE SURVEY**

A lot of research work has been done in this area from the past two decades and is still going on. Here we are putting light on some of the research work that has been done till now.

Davison focused on the idea that linked web pages have similar or related content. It says that the anchor text and the text around anchor text describe the web page which is linked to it. This idea proved to be a revolution in the area of prefetching and is being used by many researchers, crawlers and search engines today. This paper conducted an experiment or test to know whether the pages which are linked together or which are physically close in the space are also related to each other in sense of content of the web pages. This means that the pages with spatial locality, also have temporal locality. This paper explains that the designer of the web page always add links which are related to the topic of the present web page except in the case of advertisements which are used for making money. The web pages do not have random links, the links are always of the related topic.

Ibrahim made use of neural networks for making predictions for user's future web requests. The keywords of the anchor texts of the links which are embedded in the page are matched with the keyword list. Keyword list contains the keywords that occurred in anchor texts of URL's accessed by the user in its web history. If these keywords had occurred for the first time then they are added to the keyword list. Keywords are then used as input and net is calculated for each URL neuron based on the weights associated with the keywords by the predictor. Each neuron computes a weighted sum of a number of inputs and compares result with the predefined threshold value. The result is called as net. The net of the URLs are used to determine the priority of the URLs that are to be pre-fetched.

Javed I. Khan and Qingpao Tao used the webspace organization for accelerating web prefetching. They proved the fact that organization of web structure can have a great impact on the prefetching performance. Khan stressed on finding some regular structures or patterns in the web collection and he focused on exploiting such regular structures for the prediction of future web requests.

Craig E.Wills, Mikhail Mikhailov and Hao Shang proposed the idea of bundling all the embedded objects in a web page into a single unit called bundle and the whole bundle was retrieved from the server. This reduced time of making multiple connections. However, this technique involved unnecessary prefetching of web objects as all embedded objects will never be used and it also increased unnecessary load on server. This idea was further used by Alexander. Craig E.Wills, Mikhail Mikhailov and Hao Shang used the concept of pipelining to retrieve objects from the server instead of making multiple TCP connections and this idea was used by Alexander in prefetching techniques which provided positive results.

Alexander made use of object bundling for semantic link prefetching. He had proposed a technique Semantic Link Object Bundling (SLOB) prefetching which. In this method the slower web pages are bundled with the faster loading web pages. It makes use of the hyperlink structure of the collection of web pages along with semantic information associated with the link to decide which slower loading web objects are to be bundled with which faster loading web objects. The bundle was retrieved from the server instead of independent objects which helped in saving propagation time and making multiple individual connections.

Javed I. Khan and Qingpao Tao also presented a prefetching technique for reducing the response time. It used fragment streaming to minimize pre-load. Each node is divided into two parts- the lead and the stream segments. The system loads two parallel streams. In one, it pre-fetches the stream segment of current document while in other it pre-fetches the lead segment.

P. Venketesh, Dr R. Venkatesan and L. Arunprakash proposed a semantic web prefetching scheme that makes use of anchor texts present in the web page to make effective predictions. It applies Naïve Bayes classifier for computing probability of embedded URL's. The tokenizer parses the web page to extract anchor links with the associated anchor texts. When an anchor link is clicked by the user, the tokenizer moves the associated tokens into the user token repository. The user token repository has collection of tokens with their frequencies i.e. the number of times that particular token is present in the anchor link clicked by the user The probability of each anchor link is computed by applying Naïve Bayes classifier on the anchor text with reference to tokens maintained in the user token repository. The anchor links with higher probabilities are chosen and the request is sent for retrieval of respective web objects from the web server.

Sonia proposed an approach which works on the semantic preferences of the tokens present in the anchor text associated with the URL's. It makes use of semantic information which is explicitly embedded with each link to prioritize the links which are considered for further evaluation. Semantic association is computed between the tokens of the links and then the weightage is associated in order to improve the prediction accuracy. A semantic type is associated with each hyperlink using XML tags. Semantic type indicates a semantic relation between two web pages. Various semantic types are as follows : sequential(seq), similar(sim), cause-effective(ce), implication(imp), subtype(st), instance(ins), reference(ref).

- Sequential: This type signifies that the two web pages are in sequence i.e. they appear one after another in the structure of website.
- Similar: This type signifies that the two web pages are similar or they have similar content.
- Cause-effective: This type signifies that one page is cause of another.
- Implication: This type signifies that semantics of one page implies another.
- Subtype: This type indicates that one web page is subpart of another web page.
- Instance: This type indicates that one web page is instance of another web page.
- Reference: This type indicates that one web page contains the abstract view of information while another web page contains the elaborated details of that information.

The priority sequence of these semantic tags is as follows:
seq>sim>ce>imp>st>ins>ref
We are going to make use of such semantic association between the web pages for the prediction of future web requests. Out of various semantic types we are going to focus on the sequential and similar association between the web pages. Next section presents the proposed work which contains the algorithms for sequential and semantic relation between web pages.

**PROPOSED WORK**

This paper is an attempt to provide algorithms for sequential and similar relations between web pages. The simulation for both types of relations has also been done. This simulation works mostly for all types of data on the internet like images, videos, text etc. The use of anchor texts of the hyperlinks on the present web page has been made for this work. The algorithms and explanation for them is given in the next sections.

**Algorithm for prefetching sequentially related web pages**

Internet contains different types of web pages linked together. Different type of relations exists between these linked pages. Sequential web pages are one of them which are designed to be referred after one another. They often make use of anchor texts like "next", ">>" to link the sequential page. Many researchers [6][9] proved that the sequence of the web pages should be on priority when considering the links for prefetching. In many websites, it has been found that sequential pages are linked with each other using sequential numbers as anchor texts like "1", "2" and so on. We also found that anchor texts of many sequential videos are almost similar like "The Martian Part 2" will be followed by "The Martian Part 3". So we had included these patterns in our algorithm and using these common structures or patterns of the collection of web pages we had proposed this algorithm. The pseudo code for both sequential and similar algorithms is given in table 1.1.

1. User enters the keywords on the browser.
2. These keywords (or anchor texts in the case you clicked on them to visit a web page) gets stored in a String type variable called AT1 (Anchor text 1).
3. Requested web page is displayed to the user after retrieving it from the server.
4. Extract all the anchor texts and their associated URL's on the present web page one by one and the present anchor text is stored in a string type variable called AT2 (Anchor text 2).
5. If AT2 is equal to "next", ">" or ">>".
6. Compare the present webpage's parent URL with the parent URL of the web page associated with AT2 (check if they are sibling pages). If both the parent URL's are same then place this link in the prefetching list.
7. Otherwise break the anchor text (AT1) into tokens and store them in an array (arr1) and similarly break and store the tokens of AT2 in an array (arr2).
8. Compare the lengths of "arr1" with "arr2". If length is same then continue otherwise skip all these steps and go to similar algorithm for checking similar relation.
9. Match "arr1" with "arr2", token by token (arr1[0] with arr2[0], arr1[1] with arr2[1]). If match occurs then continue matching. Otherwise,
10. Check if "arr1[x]" (present token of arr1) is a number like "1", "2" or "one", "two" etc. If yes, check "arr2[x]" if it is the next number or sequential number ( if arr1[x] is "2'" and "arr2[x]" is "3") then place this link in the prefetching list otherwise check for similar relatedness.
11. If at any point of time user requests for a web page, the prefetching cache is first checked for the requested page. If the required web page is present in the prefetching cache, the page is displayed to the user otherwise a request is sent to the server and the algorithm starts again from step 1.

**Table. 1.** Pseudo code for sequential and similar algorithms

```
Sequential()
{
  // if at any point of time user requests a new webpage GOTO Sequential()
    AT1= "keywords entered by user";
    String r1[] = AT1.split();          //break into tokens
    for( link:links)   //a loop which takes anchor texts on current webpage
//one by one
    {
      AT2= link.text();   //anchor text of present chosen hyperlink in the
//loop.
      String r2[] = AT2.split();
      If(r2.length<=3)
      {
        for( i=0 to (r2.length-1))
        {
          If( (r2[i].equalsignoreCase("next")||(r2[i]== ">>")||(r2[i]== ">"))
          {
             If(Parenturl1.equalsIgnorecase(Parenturl2))    //compare parent
//url's of present webpage &  associated hyperlink
             {//add to prefetching list}
          }
        }
      }
    if(r1.length==r2.length)
    {
      If(Parenturl1.equalsIgnorecase(Parenturl2))
      {
       for( i=0 to r1.length)
       {
        If(r1[i]==r2[i])
        {
        }
        else
        {
          switch(r1[i])
          {
           case "1" : if ((r2[i]== "2")|| (r2[i].equalsIgnoreCase ("two")))
{ //Add to prefetching list}
           case "2" : if ((r2[i]== "3")|| (r2[i].equalsIgnoreCase("three")))
{ //Add to prefetching list}
           case "3" : if ((r2[i]== "4")|| (r2[i].equalsIgnoreCase("four")))
{ //Add to prefetching list}
           case "one" : if(r2[i]== "2")|| (r2[i].equalsIgnoreCase("two")))
{ //Add to prefetching list}
           case "two" : if(r2[i]== "3")|| (r2[i].equalsIgnoreCase("three")))
{ //Add to prefetching list}
           case "three" : if(r2[i]== "4")|| (r2[i].equalsIgnoreCase("four")))
{ //Add to prefetching list}
            &…… so on
            default:
                // GOTO similar()
          }
         }
       }
     }
       else
       {
         // GOTO similar()
       }
     }
   }
}
```

```
Similar()
{
//if at any point of time user requests a web page
//GOTO Sequential()
// remove stopwords,
//  perform stemming,
// sort  elements in r1[] & r2[]
// & store in r3[] & r4[].
// Build a matrix a[m][n]  with elements of r3[]
on one //axis & r4[] on other,
// where m= row number & n= column number &
// a[i][j] contains the similarity score between
row i & //column j.
   for (k=0 to (r3.size-1))
   {
     float max[k]= r[k][1];
     for (x=0 to (r4.size()-1))
     {
       if (a[k][x]>max[k])
       {
         max[k]= a[k][x];
       }
     }
   }
  float total=0;
    for ( i=0 to max.size())
    {
     total=total+max[i];
    }
    int big;
    if(r1.size()>=r2.size())
    {
     big= r1.size();
    }
    else
    {
     big=r2.size();
    }
    float probability= total/big;
    if(probability>=0.7)
    {
      //Add to prefetching list
    }
}
```

**Algorithm for prefetching similar web page**

If sequential pages are not found, then look for similar pages. For this the use of anchor texts of the hyperlinks on the present webpages has been made again. The content of many websites had shown that similar web pages have similar anchor texts. Semantic similarity between the keywords provided by the user (AT1) and the anchor text of the hyperlinks (AT2) on the present web page is computed. Other descriptive texts like titles, text around anchor texts can also be used for this purpose. The anchor text is broken into tokens and then semantic similarity is computed between tokens of user's keyword and then by using this, semantic similarity is computed between the whole sentence (anchor text).

To compute similarity between two words, use of a method called WS4J (WordNet similarity for java), in which the use of dice coefficient formula and ontologies has been made. Dice Coefficient formula is given below.

$$\frac{2|X \cap Y|}{|X| + |Y|}$$

X, Y represents the two words between which similarity is to be computed.
$|X|$ = distance of word 'X' from the root node in the ontology.
$|Y|$ = distance of word 'Y' from the root node in the ontology.
$|X \cap Y|$ = distance of the common parent node from the root node in the ontology.

Due to lack of resources, semantic similarity between words had already been computed and had been stored in a table. For computing similarity between two sentences an algorithm is presented which is as follows:

    i. Remove stop words from "arr1" and "arr2".
    ii. Perform stemming on all the tokens in "arr1" and "arr2" to bring them back into their root form.
    iii. Perform sorting on "arr1" and store tokens in "arr3" and after sorting "arr2", store tokens in "arr4".
    iv. Build a matrix R[m,n], where R[x, y] is the semantic similarity between the words, at position 'x' in arr3 and position 'y' in arr4.
    v. Find semantic similarity using WS4J dice coefficient method between each pair of tokens.
    vi. Find maximum of all the semantic similarities in each row, & store them in an array named 'MAX'.
    vii. Add all the elements of array 'MAX' and store it in variable named 'TOTAL'.
    viii. Find the length of bigger array between arr3 & arr4.
    ix. Divide 'TOTAL' by length of bigger array to give required probability.
    x. Compare the result with a predefined threshold value.
    xi. If it is greater than threshold, then that link is placed in prefetching list.

**Examples**

This section provides examples of both sequential and similar types of relation.

*Example 1: Sequential type*
1. The keywords say "HTML tutorials" are entered in the browser.
2. These keywords are stored by the search engine & named as 'AT1'.
3. The user clicks on any of the link provided by the user and a web page gets displayed.
4. Anchor texts of all the links on the present web page are retrieved one by one and the present anchor text is stored in variable 'AT2'.
5. Break the anchor text of AT1 and AT2 and store them in arrays 'arr1' and 'arr2' respectively.
6. Check if 'arr2' contains tokens like *"next", ">>", ">"* etc.
7. If any of these tokens are present then compare the parent URL of present web page *(http://www.w3schools.com/html)* with the parent URL of web page associated with AT2. *(http://www.w3schools.com/html)*.
8. Here, the match occurs, so this link is placed in the prefetching list, otherwise

*(Now say AT1 was "The Martian part 1" and AT2 was "The Martian part 2").*

9. Compare the lengths of 'arr1' with 'arr2'. If match occurs then continue, else skip all these steps and go to similar algorithm to check similar relation.
10. Now match occurs, then start matching 'arr1' with 'arr2' token by token like arr1[0]*("The")* with arr2[0] *("The")* and arr1[1]*("Martian")* with arr2[1] *("Martian")* and so on.
11. If match occurs for arr1[i] with arr2[i], then increment 'i' and continue matching till lengths of the array. If it doesn't then check if arr1 contains a token like '1', '2' or *"one" , "two"* etc.
12. If yes, then check if arr2 contains the next number in sequence, (here arr1[3] contains '1' and arr2[3] contains '2').

13. If yes, match their parent URL's.
14. If URLs match, then place this URL in the prefetching list. *(This will also work for sequential pages like if on page '1', automatically pre-fetch page '2')* , else go to similar algorithm.
15. (If at any point of time user makes a request for a web page, prefetching cache is first checked for the required web page. If the required web page is present in the prefetching cache, it is displayed to the user otherwise a request is sent to the server. As soon as new web page is displayed to the user, the algorithm for sequential starts from step 1)

*Example 2: Similar pages*

    AT1 = what is the best books on operating system
    AT2 = what is a good book to learn fundamentals of computer operating system

I. Remove stop words from arr1, arr2.
II. Perform stemming on all the words in both the arrays. *(To bring back the words into their root form).*
III. Sort arr1 and arr2 and store them in arr3 & arr4.
    arr3= [book, good, operating system]
    arr4= [book, computer, fundamental, good, learn, operating system]
IV. Now build a matrix R[m,n] where m= length of array 'arr3' and n= length of array 'arr4'. R[x, y] is the semantic similarity between the words, at position 'x' in arr3 and position 'y' in arr4. A matrix table for this example is given in table 1.2.
V. Find the maximum similarity in every row, shown in 'MAX'.

**Table. 2.** Matrix table for example 2

|  | **Book** | **computer** | **fundamental** | **good** | **learn** | **operating system** | **MAX** |
|---|---|---|---|---|---|---|---|
| **Book** | 1 | 0.1 | 0 | 0.1 | 0 | 0.1 | **1** |
| **Good** | 0.1 | 0.1 | 0 | 1 | 0 | 0 | **1** |
| **operating system** | 0.1 | 0.2 | 0 | 0 | 0 | 1 | **1** |

VI. Find the sum of all the elements under 'MAX' and name it 'TOTAL'.
    TOTAL= 1+1+1
        = 3.0
VII. Choose the length of the biggest array out of two *(arr3=3, arr4=6, biggest=6)*.
VIII. Divide 'TOTAL' by the length of biggest array to find the probability.
    Probability= 3.0/6
        =0.5
IX. If at any point of time user requests for a web page, the prefetching cache is first checked for the requested page. If it exists, the required page is displayed to the user otherwise a request is sent to the user and we go to step 1.

*Example 3: Similar pages*

    AT1 = what is the best books on operating system
    AT2= what is a good, complete book on operating system
    arr3= [book, good, operating system]
    arr4= [book, complete, good, operating system]

A matrix table for this example is shown in table 1.3.

**Table. 3.** Matrix table for example 3

|  | **Book** | **complete** | **Good** | **operating system** | **SUM** |
|---|---|---|---|---|---|
| **Book** | *1* | *0* | *0.1* | *0.1* | *1* |
| **Good** | *0.1* | *0* | *1* | *0* | *1* |
| **operating system** | *0.1* | *0* | *0* | *1* | *1* |

TOTAL=1+1+1
=3.0
Probability= 3.0/4
= 0.75

The links which are above or equal to predefined threshold (0.7 here) are sent to the prefetching list. Webpages associated with links in the prefetching list, are pre-fetched and stored in the prefetching cache. When the user requests a new web page, prefetching list is first checked for the required web page. If the required webpage is present in the prefetching cache it is displayed to the user otherwise a request is sent to the server.

**CONCLUSION**

This paper focuses on the semantic relation between web pages which is used to improve performance of web prefetching. Out of these (sequential, similar, cause-effective, implication, subtype, instance, referential) semantic relations, this paper provides algorithm for sequential and similar relations. The sequential type has the highest priority followed by similar type. Earlier researchers had implemented these associations using semantic tags which were embedded in the web page during their designing. For adding this semantic information, XML tags were used. However it is very unlikely that all the designers will care to add such information while designing. This thing will take some time to get implemented. Moreover the user's system has to depend on the server who is going to provide this information to the client along with the required page. So, this paper is an attempt to provide algorithms and simulations for sequential and similar semantic relatedness between web pages which are implemented on the client-side.
It helps in better prediction of future web pages as compared to prefetching methods which were solely based on user's history. Plus, this method is not dependent on the designers for the addition of semantic information with the help of XML tags during web page designing. It works independently on the client side. Next section focuses on the possible future work in this area.

**FUTURE WORK**

This paper presents algorithms and their simulation for implementing sequential and similar relatedness between web pages. In the future, some more common structures of the webpages will be used for improving the existing algorithms. Future work is going to focus on the algorithms for implementing remaining types of semantic relatedness between web pages which are cause-effective, instance, subtype, implication and reference. Due to lack of resources, the use of ontologies could not be made to compute similarity between tokens. If the resources become available in the future, the full implementation of these algorithms will be done.